# Comparison of two models of electric neuro-stimulation and consequences for the design of retinal prostheses

Erich W. Schmid  (Institute for Theoretical Physics, Univ. Tuebingen, Germany)

Two simple mathematical models of electric neuro-stimulation are derived and discussed. It is found that the common injected-charge model is less realistic than a model, in which a latency period, which follows after a short electric pulse, plays a role as important as the electric pulse. A stimulation signal is proposed that takes advantage of these findings and calls for experimental testing.

## 1. Introduction

The idea of building a retinal prosthesis that can, at least to some extent, restore vision to a blind person, has fascinated people since decades [1-5]. Much has been achieved in micro-electronic technology, in our understanding of biocompatibility and biostability of implanted materials and in surgical approaches; excellent overviews can be found at Ref[6-11]. A central part of research has been, and still is, the electric stimulation of neural tissue of the retina. In the present paper we focus our attention on this and search for design parameters that can be optimized.

Since we will use some basic concepts of electric stimulation, we will give a very brief account in Sections 2 and 2.1. In Sect. 2.2 we will discuss the common practice of considering the injected charge as the key quantity responsible for stimulation. We will check how well this assumption agrees with experimental data. We then proceed in Sect. 2.3 to a model that emphasizes latency rather than injected charge. We'll find that this model agrees quite well with known experimental findings.

In Sect. 3 we will look for design parameters that we might use to optimize electric stimulation. We will concentrate on consequences following from the model presented in Sect. 2.3. We will also briefly talk about parameters following from the fact that the electric current field that transmits the stimulation signal is a vector field and not a scalar field, and that its strength depends on space and time.

## 2. The models

In a retinal prosthesis we can clearly distinguish between two functional entities. The first entity can be regarded as a transducer that converts the image to be seen (i.e. optical information) into biological information (i.e. electric signals). It may consist of a camera plus an electronic processing unit outside of the eye, or it may sit inside the eye in form of a microchip with photo-diodes plus a processing unit. This first entity will not be discussed in this paper. The second entity is the interface between the artificial vision device and the natural visual system. This interface must transmit the electrical information into the visual pathway. Today this is generally achieved by electric stimulation of neuronal tissue. The electric stimulation is performed by an electrode array on a silicon chip or on a carrier substrate like polyimide. In the epiretinal approach the transduction is usually performed entirely outside the body using a camera and microcomputer. The electrical information is then transferred wirelessly to the eye, where an electrode-array is positioned at the surface of the inner retina, facing the ganglion cell layer. In the subretinal approach the chip responsible for transduction and stimulation is located between the (degenerated) photoreceptor-layer and pigment epithelium/choroid, with electrodes facing the retina and vitreous. Stimulation signals are desired to interface with the visual pathway at its first neurons, namely at bipolar cells, horizontal cells or amacrine cells.

The physical quantity responsible for the communication between electrodes and nerve cells is an electric current in the neuronal tissue, which is caused by a voltage applied between electrode(s) and counter electrode(s). By Ohm's law the current field has an associated electric field. The electric field can enter into a nerve cell and modify its trans-membrane potential. The latter process is understood since the pioneer work of Hodgkin and Huxley[12,13] and is mathematically described by the cable equation[14,15].

Or present interest is the time-dependence of the stimulation current and its effect on a nerve cell. We'll compare two very simple mathematical models and we'll try to learn something about the design parameters of electric stimulation signals.

**2.1. Some basic concepts**

A stimulation signal can be produced either by a current generator or by a voltage generator. In recent applications, voltage generators have been used[16]. As shown in Ref[17] both monophasic rectangular and biphasic rectangular voltage pulses have been tested. In Fig. 1 we show a monophasic rectangular voltage pulse and the induced current. Such pulses, with a duration of 0.5 milliseconds are used in the implant described in Ref[16]. The current rises sharply when the voltage is turned on, and then decays more or less like an exponential function. Note, that the voltage drop from $U=V$ to $U=0$ at $t=T$ means that the voltage generator is shortcutting the electro-chemical cell. A discharging current arises with a shape similar to the one of the charging current, but with opposite sign.

The question is how, in detail, does a nerve cell react to an electric current in the extra cellular space? As has been said before, the pioneer work of Hodgkin and Huxley[12,13] gives us some understanding. And H.C. Tuckwell explains in his text book[15] how the old cable equation by Heaviside[14] describes the electrochemical processes in nerve cells. The manifold of solutions of the cable equation[18] tells that the sharp rise at the beginning of the time evolution of the current is the main cause for the signal to enter into the nerve cell, thereby changing the trans-membrane potential and initiating the generation of action potentials.

Somehow, experimental experience of the early days suggested another view of the mechanism of electro stimulation, namely that injected charge, which is the time integral over the current, causes stimulation[8,19]. It is indeed intriguing to think that charge can be injected like paint of a spray can. After all, it is charge on the outer surface of the cell membrane that attracts counter charge on the inner surface and modifies cell polarization. In the present paper we want to challenge this point of view, and present another simple view, instead. Let us first take a look at the injected charge model, and then present the other model.

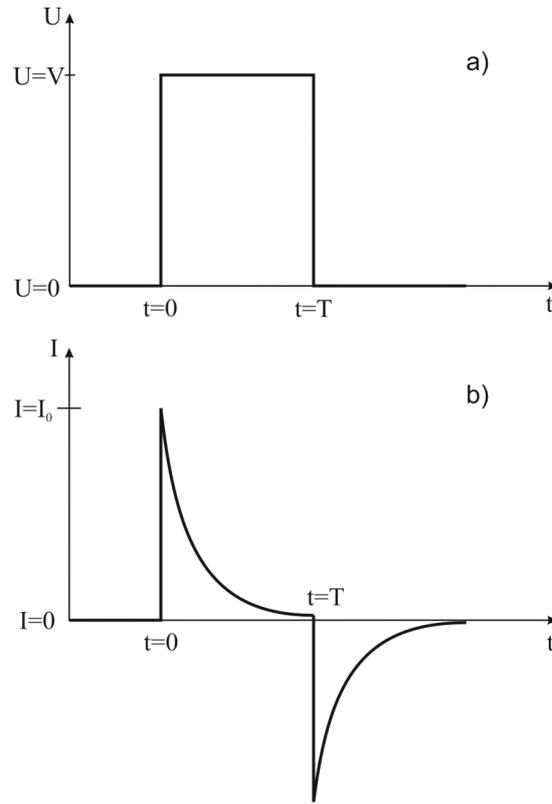

**Figure 1** Voltage *U(t)* and electric current *I(t)* of a monophasic stimulation pulse. a) Rectangular voltage pulse of height *V* and duration *T*; b) Electric current through the electrochemical cell formed by electrode-tissue-counterelectrode.

We should emphasize that, for both models, we make the simplest possible, nontrivial approximations. The reason for simplicity is, that refinements introduce more fitting parameters. And with many fitting parameters it becomes too easy to fit experimental data. On the other hand, we should be aware that simplifications invite criticism; but criticism is always welcome.

**2.2. The injected charge model**

We want to calculate the injected charge between *t=0* and *t=T*. Integration becomes easy if we let the current *I(t)* start at $I=I_0$ and describe its decay by a single exponential function, with decay constant $\kappa$,

$$I(t) = I_0 \exp(-\kappa t). \tag{1}$$

The injected charge *Q(T)* as a function of pulse duration *T* is obtained by integration,

$$Q(T) = \frac{I_0}{\kappa}(1-\exp(-\kappa T)). \tag{2}$$

We make another approximation and say that the initial current $I_0$ is proportional to the applied voltage *V*,

$$I_0 = cV. \tag{3}$$

Inserting this we get

$$Q(T) = \frac{cV}{\kappa}(1-\exp(-\kappa T)). \tag{4}$$

Now we demand that there is a threshold value $Q_{thres}$ of the injected charge $Q(T)$, for which the stimulation signal is strong enough to elicit a visual sensation. The key idea of the injected charge model is that this threshold charge is a constant, i.e. it is independent of $V$ and $T$. If we put $Q(T)$ at its threshold value $Q_{thres}$ we obtain the threshold voltage $V_{thres}$,

$$Q_{thres} = \frac{cV_{thres}}{\kappa}(1 - \exp(-\kappa T)). \tag{5}$$

We are interested in getting the threshold value $V_{thres}$ of $V$, for given pulse duration $T$. Therefore, we resolve for $V_{thres}$ and get

$$V_{thres} = \frac{\kappa Q_{thres}}{c} \frac{1}{1 - \exp(-\kappa T)}. \tag{6}$$

This is the fitting equation of our simplified injected charge model. It is a two-parameter formula. Clearly, for $T \rightarrow \infty$ we get $V_{thres} = (\kappa Q_{thres})/c$; therefore, $((\kappa Q_{thres})/c)$ is the first parameter of the model, giving us the asymptotic value of the threshold voltage $V_{thres}$. The second parameter is $\kappa$, the exponential decay rate of the current. The fitting formula (6) will yield a fairly good fit to our experimental data, as will be seen shortly.

**2.3. The model based on electric pulse and latency**

Let us look once more at Fig. 1 b and see what might be inadequate with the injected charge model. We are interested in a voltage threshold to elicit a visual sensation. Turning on the voltage will lead to a current peak, as seen in Fig. 1 b. This current peak will produce, by Ohm's law, an electric field in the tissue. It is this field or, mainly its time-derivative, which is responsible for a change of the trans-membrane potential of adjacent neural cells. With increasing time the current decreases and with it the stimulating electric field and the time derivative of the electric field. At $t = 1/\kappa$, about half ways between $t = 0$ and $t = T$ in our figure, only about *37 %* of the current is left and, at twice that time, only about *14 %* is left. We doubt that such small currents and, consequently, such small electric fields have any sizable biological effect. From experiment we know that going from a pulse duration of $T = 0.5$ milliseconds to $T = 1$, or *2*, or *4* milliseconds yields lower thresholds. We believe, however, that this is not due to the increase of injected charge. We hypothesize that it results from leaving the nerve cell unaffected before shortcutting the electro-chemical cell and inducing an electric field of opposite direction.

We want to cast this hypothesis into a mathematical model. We consider electrostimulation to be a two-step procedure: (1) an electric impulse of duration $\Theta$, (2) a latency period of duration $T$. For illustration we modify Fig. 1, as shown in Fig. 2. At time $t = \Theta$ a circuit breaker turns off the charging current of the electrochemical cell and disconnects the voltage generator. The neural tissue, which has been exposed to the electric pulse, is now free up to $t = T$. The changes in trans-membrane potential induced by the pulse will eventually lead to modulation of neurotransmitter release after a latency period. During this latency period T no additional pulse and no short-cutting of the electro-chemical cell will influence this process.

Again, we want to describe stimulation by a very simple model that will yield a fitting equation for threshold voltage versa latency time, with only two fitting parameters.

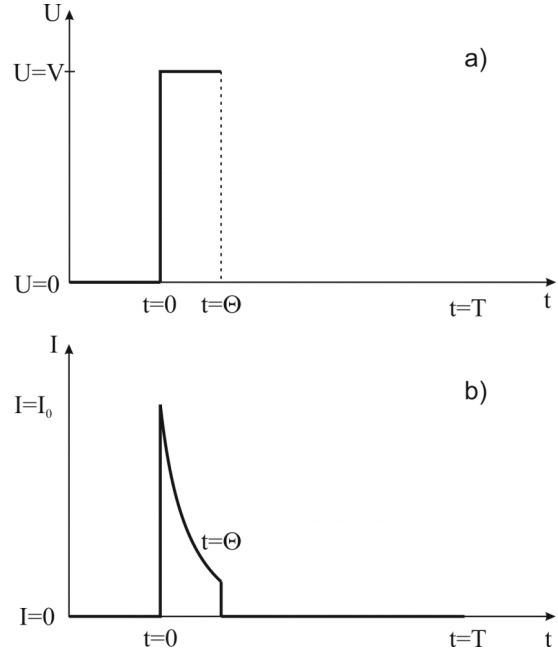

**Figure 2** Voltage and electric current, for monophasic stimulation by a short electric pulse plus latency. a) The voltage generator is disconnected at $t=\Theta$; b) the current is turned off by the circuit breaker and remains equal to zero till the end of the latency period at $t=T$.

First we define a strength function $S(V,\Theta,T,...)$. The value of this function tells whether a signal is strong enough to elicit a retinal response. Its threshold value $S_{thres}$ will be our main concern. The two-step model suggests that this strength function factorizes into a product of an amplitude $A(V,\Theta,...)$ and a function $f(T)$,

$$S(V,\Theta,T,...) = A(V,\Theta,...) f(T). \tag{7}$$

Both of these functions are fitting functions. They can, in principle, be determined by experiment; in practice they are unknown functions. We expand the amplitude $A(V,\Theta,...)$ into a Taylor series in $V$, and the function $f(T)$ into a Taylor series in $T$. We truncate both series after the first non-trivial term, which is the linear one,

$$A(V,\Theta,...) = a + b V, \qquad f(T) = \alpha + \beta T. \tag{8}$$

Since we expect $f(T)$ to begin at $f(0) = 0$ we may put $\alpha = 0$. We then get

$$S(V,\Theta,T,...) = (a \beta + b \beta V) T. \tag{9}$$

Again, we are interested in the threshold strength $S_{thres}$ and the corresponding threshold voltage $V_{thres}$, while all other parameters like $\Theta$ and $T$ are kept fixed,

$$S_{thres}(V_{thres},\Theta,T,...) = (a \beta + b \beta V_{thres}) T. \tag{10}$$

As we did before with $Q_{thres}$, we assume that the threshold value $S_{thres}$ of the strength function $S$ is independent of $V_{thres}$ and $T$. Resolving for $V_{thres}$ we then get our fitting formula,

$$V_{thres} = -\frac{a}{b} + \frac{S_{thres}}{b\beta} \frac{1}{T}, \tag{11}$$

or,

$$V_{thres} = \frac{(-a)}{b}(1 + \frac{S_{thres}}{(-a)\beta}\frac{1}{T}). \tag{12}$$

This fitting formula is well known. If we rename the fitting parameters to *(-a)/b =R* (rheobase) and $S_{thres}/(-a\beta)=C$ (chronaxie) we get the classical strength-duration fitting formula,

$$V_{thres} = R(1 + \frac{C}{T}). \tag{13}$$

Let us see now how well the two models fit experiment. Experimental data we take from Fig. 6 of Ref.[17] and present them in Fig. 3. This figure shows the perception threshold voltage as a function of the duration *T* of a monophasic rectangular voltage pulse. Averages are taken from 4 single experiments performed in one human test person; error bars are the one-sigma deviation.

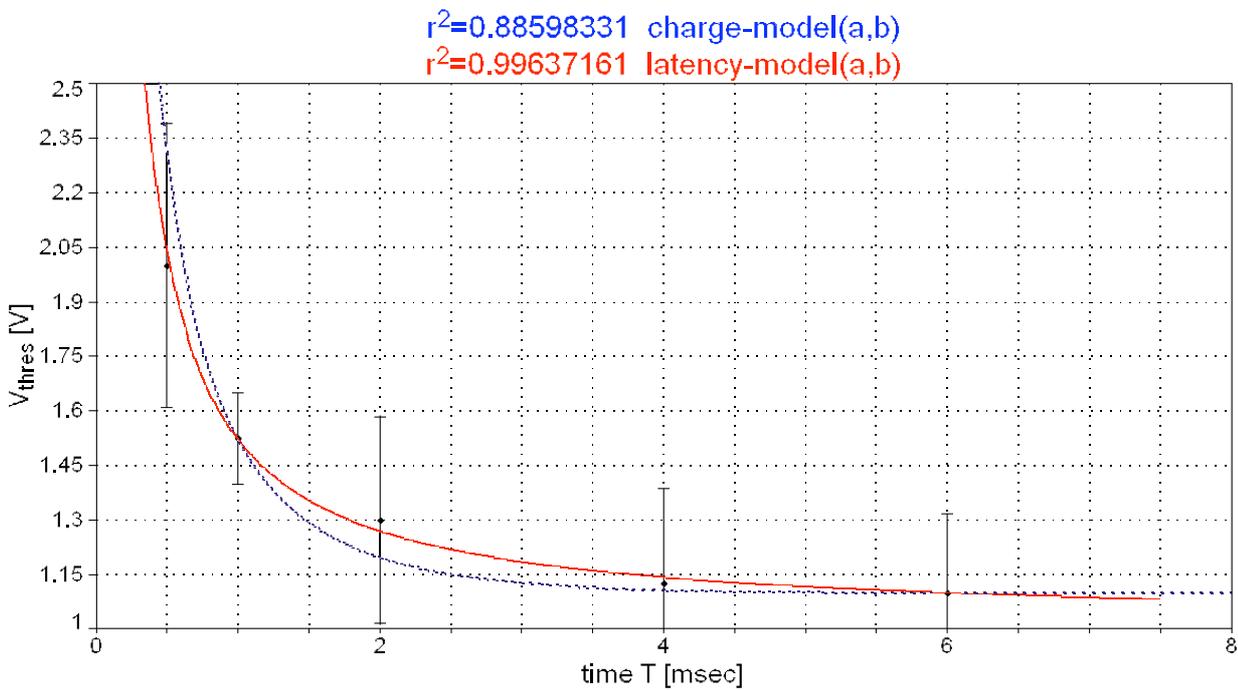

**Figure 3** Comparison of the two mathematical models fitted to experimental data. Within this experiment the voltage threshold to elicit a visual sensation by a rectangular voltage pulse with variable pulse duration (T) has been determined. Plotted are the mean values of 4 experiments with +/- 1SD. The dotted blue line indicates the results using the charge model, the solid red line using the latency model.

It would be easy to do the mouse click on the computer in order to get a least-squares fit, for both of our models. For better insight we do something else. We do not know the asymptotic value of our threshold voltage from measurement. But *T = 6* milliseconds is a rather long time, almost asymptotic. So we take this point and fit it exactly; the same has been done in Ref[17]. At *T = 2* milliseconds we find a point with a small error bar. This point would dominate in a weighted least squares fit. So we fit also this point exactly. The so obtained fitting curves are shown in Fig. 3. The dotted blue line shows what we get from the injected-charge model; the solid red line shows what we get from the latency model, eq. (13).

Comparing the two models we see that we get a better fit from the latency model. But let's be fair and say that this fact is not significant. Our emphasis was on simplicity;

we wanted only two fitting parameters, and we accepted simplifications. Therefore, none of the fits should be perfect. However, the data do not indicate that the injected-charge model is any better than the latency model.

### 3. Implications on the design of retinal prostheses

Much has been achieved in the field of electric stimulation of nerve cells. But, for satisfactory imaging we need smaller phosphenes, and we should be able to transmit a grey scale. Are there still possibilities left to improve electric stimulation?

Let's try to get some order into our effort of searching for new possibilities. The stimulation signal is transmitted via an electric current that flows through the tissue. The current density is a vector field $\vec{j}(x,y,z,t)$. Actually, this vector field is present only in the extracellular saline. At the moment we don't want to go that far. We accept the approximation of ignoring the fine structure of the tissue and work with an averaged current field.

The current density $\vec{j}(x,y,z,t)$ is a function of time $t$ and of space $x,y,z$. And it is a vector, which means that the current has a direction. This suggests an order. To be able to come to better stimulation paradigm we have to ask: (1) what can be done to optimize *time* dependence? (2) How can we optimize *space* dependence? (3) Can we gain something by modifying the *direction* of the current? We'll deal with space dependence and dependence on the direction of the current in forthcoming papers. Here, we want to present a proposal about time dependence, which is meant to hold for unipolar, bipolar and multipolar electrode arrays.

Much experience has been gained with monophasic and biphasic rectangular voltage pulses; a monophasic rectangular voltage pulse is shown in Fig. 1 a. When a second rectangular pulse of opposite sign follows, immediately after the first pulse, one calls it a biphasic voltage pulse. It is called "anodic first" when $V$ is positive and "cathodic first" when $V$ is negative.

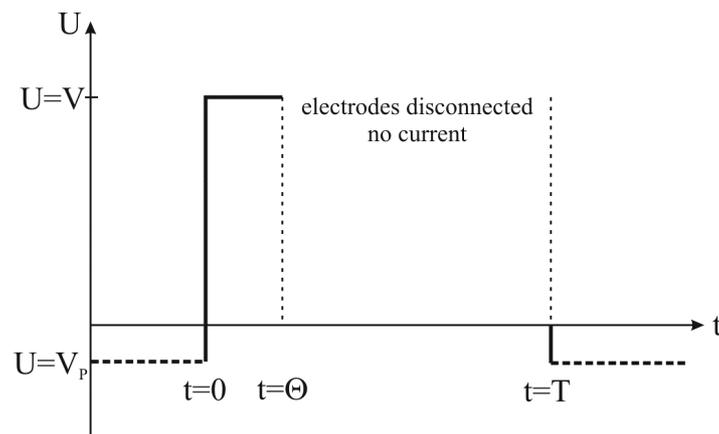

**Figure 4** Proposed shape of a voltage pulse, with stimulation voltage $V$ and preset voltage $V_p$. Solid line: stimulating voltage. Broken lines: preset or reset with limited current and voltage, in order to avoid damage and to avoid starting a new latency period. The circuit is open between $t=\Theta$ and $t=T$.

The new time characteristic of the stimulating signal is shown in Fig. 4. Before we start to transmit the stimulation signal we reset, or preset, the electrochemical cell formed by electrode(s)–tissue–counterelectrode(s). Reset means voltage equal to zero.

Preset means voltage equal to some small below-threshold value $V_p$. The sign of $V_p$ is opposite to the one of $V$, because $V_p$ serves to increase the voltage jump. The voltage $V$ is turned on at $t=0$. It is switched off at $t=\Theta$. A circuit breaker guarantees that the electrochemical cell floats between $t=\Theta$ and $t=T$. During that period we expect some slow and complicated self-discharging process. The Helmholtz layers of electrode and counter electrode will be completely discharged (and then charged to $V_p$) in a new reset (or preset) period. At the end of the last signal of a series of signals, the preset must be a reset, i.e. $V_p$ must be put equal to zero. A resistor, or other electronic device, has to guarantee that preset or reset is done gently, without any peak current that could eventually kick off a new latency period.

The advantage of the new time-dependence of the stimulation signal is that it offers one more degree of freedom for design optimization, namely $\Theta$, in addition to the old degrees of freedom $V$ and $T$. For short $\Theta$ the restrictions on the voltage $V$ can be weakened. As is seen from Fig. 3, for instance, ± 2 Volt is considered a safe limit to avoid tissue damage and electrode damage, for a given electrode material and geometry. This limitation is valid for direct current and thus allows pulse durations of several milliseconds. For high-frequency alternating current ($10^4 - 10^5 Hz$) much higher voltages are still safe. This is true because chemical reactions, like the formation of hydrogen molecules and oxygen molecules, are not instantaneous and need time. We conclude that microsecond pulses, instead of millisecond pulses, allow for safe application of voltage pulses that are much higher than ±2 Volt. Other than in electrochemical experiments, with pure electrolytes and clean electrode-electrolyte interfaces, we are working with a biological system. Higher voltages may help to overcome biological obstructions at the interface. There is another advantage of short $\Theta$ when the array of electrodes is unipolar and a sequential firing order is employed to avoid collective field effects, because short $\Theta$ increases the time gaps needed for sequential firing.

We want to enter into the visual path at its beginning, right after the cone or rod cells, which are usually lost in the process of retinal degeneration. We want to stimulate bipolar cells or even amacrine and horizontal cells. We do not want to stimulate ganglion cells. The physical function of a bipolar cell is to integrate very small signals; a long latency is acceptable in order to enhance sensitivity by loosing some temporal resolution[20]. We can adapt to such a behavior by choosing a small $\Theta$ and a large T. This would then be something like an "address" of bipolar cells. Similar thoughts have been presented in Ref.[8]; bypassing the ganglion cell bodies and their axons, and addressing bipolar cells is important for both epiretinal and subretinal applications.

There might be the question of whether a step function at $t=0$ in Fig. 4 is a good choice. From solving the cable equation [18] we know that a large time derivative of the current is favorable for signal penetration into the nerve cell. For that reason we don't think that smoother functions at $t = 0$ are more favorable than a step function.

Would it be difficult to implement the desired time-dependence of the signal in known devices? In the device described in Ref [16] the pulse duration is determined by the duration of the power supply to the amplifiers. It seems to be straightforward to let the power supply start at time $t=0$ and stop at time $t=\Theta$. In Ref[16] the pull-down circuit becomes active at the moment when the power supply stops. Here, one would have to install a circuit which causes a delay from $t=\Theta$ to $t=T$.

## 4. Conclusion

A strength-duration curve had been measured with several human blind volunteers and has been presented in Ref[17, 21, 22]. The experimental data are well explained by a model that uses as critical quantity the injected charge, i.e. the time integral over the electric current. It is shown in this paper that another model explains the data just as well, or even better. In the latter model the current is cut off after its initial peak and a latency period comes into play.

Based on the second model we propose an electric stimulation signal that is a combination of voltage control and current control. The initial current peak is voltage controlled, for the reason of not damaging the tissue and avoiding corrosion of electrodes. After the current peak has passed, current control causes the current to become zero, until a latency period comes to an end. After that the electrodes are either discharged under current control, or preset for another stimulation cycle.

The present proposal relies on theoretical considerations and some preliminary experimental findings. Therefore we are eager to learn from future experimental studies, which may prove or discard these considerations.

## Acknowledgement
Collaboration of Dr. R. Wilke is gratefully acknowledged.